\documentclass[12pt]{iopart}
\usepackage{iopams}     
\usepackage{graphicx}
\usepackage{epsfig}
\usepackage{color}
\usepackage{url}
\usepackage{times}
\usepackage{bm}         
\usepackage{times}

\begin{document}

\review{Numerical relativity confronts compact neutron star binaries:
a review and status report
}

\author{Matthew D. Duez}

\address{$^1$ Center for Radiophysics and Space
    Research, Cornell University, Ithaca, New York, 14853}

\begin{abstract}
We review the current status of attempts to numerically model
the merger of neutron star-neutron star (NSNS) and black hole-neutron
star (BHNS) binary systems, and we describe the understanding of
such events that is emerging
from these calculations.  To accurately model the physics of NSNS and BHNS mergers
is a difficult task.  It requires solving Einstein's equations
for dynamic spacetimes containing black holes.  It also requires evolving the
hot, supernuclear-density neutron star matter together with the
magnetic and radiation fields that can influence the post-merger dynamics. 
Older studies concentrated on either one or the other
of these challenges, but now efforts are being made to model both relativity
and microphysics accurately together.  These NSNS and BHNS simulations are
then used to characterize the gravitational wave signals of such events
and to address their potential for generating short-duration gamma ray bursts.
\end{abstract}


Compact object binaries consisting of either two neutron stars (NSNS binaries)
or a black hole (BH) and a neutron star (NS) (BHNS binaries) are unique
laboratories for studying both strongly-curved spacetime and
supernuclear-density matter.  Such binaries radiate away orbital energy
through gravitational waves, causing the two objects to slowly spiral in closer
to one another and, eventually, to collide and merge.

Estimating the rate of NSNS/BHNS mergers is not easy.  In our galaxy, there
are several observed NSNS binaries that should merge due to gravitational
radiation-induced inspiral in less than a Hubble time.  (The double pulsar
system J0737-3039 will coalesce in 85Myr~\cite{Burgay:2003jj}.)  Extrapolating
from these observations, the rate of NSNS mergers per Milky Way equivalent
galaxy has been estimated in the range 40-700 per Myr~\cite{Kalogera:2003tn}. 
In contrast, no BHNS binaries have yet been detected. 
The merger rate can also be estimated theoretically. 
NSNS and BHNS binaries are thought to form from binary systems of two
massive ($>8M_{\odot}$) stars.  Population synthesis calculations model
the evolution of a large number of massive stellar binaries to determine the
likelihood of producing tight NSNS or BHNS systems (see
e.g.~\cite{Popov:2004gw,11767} for details.). There are
large uncertainties regarding several aspects of the evolution of these
binaries,
leading to more than an order of magnitude
uncertainty in the predicted NSNS and BHNS merger rates.  Nevertheless, the
predicted NSNS merger rate is consistent with that inferred from
the number of observed systems, and the BHNS rate is predicted to be about two orders of
magnitude lower~\cite{Belczynski:2006zi}.

NSNS/BHNS mergers might have a number of observable effects. 
The most direct way to observe them is through the
gravitational waves emitted during the late inspiral and merger.  Population
synthesis calculations predict that Advanced LIGO should observe of order 10
NSNS merger events per year, and about one BHNS event~\cite{Belczynski:2006zi}. 
These events are also thought to be the most promising explanation of the
majority of observed short-duration gamma ray bursts (GRBs).  Most GRB models
require a BH surrounded by a large accretion disk.  Energy is extracted,
either from the disk or the black hole's spin, and is dumped into a baryon-poor
region around the poles.  This drives the ultrarelativistic outflow needed to
explain the observed athermal spectra of GRBs.  (For a review of short GRB
progenitor models, see~\cite{Lee:2007js}.)  Only numerical simulations can
test the assumptions of this model:  a massive post-merger accretion disk, a
baryon-poor region, and efficient energy extraction through neutrinos or magnetic
fields.  Finally, NSNS/BHNS mergers may be important for understanding the observed
abundances of the heavy elements that are formed by rapid neutron capture
in the r-process~\cite{1974ApJ...192L.145L}.  (See~\cite{2007PhR...450...97A}
for a recent review of
the r-process and its proposed sites.)  Here again simulations are needed
to determine how much and what kind of matter is ejected during such mergers.

In the next section, we review the physics and numerics involved in modeling
NSNS/BHNS mergers.  Then, for each type of binary, we describe the current
understanding of the merger process derived from simulations.  For another
useful review of this field, see Faber~\cite{2009CQGra..26k4004F}.


\section{The numerical challenge}
\label{challenge}
Neutron stars have compactions in the range $GM/Rc^2\approx 0.1-0.2$, so we
expect general relativistic (GR) effects to be important to their evolution. 
Important studies have nevertheless been carried out using Newtonian physics
(see below).  For BHNS binaries, this means the BH is modeled as a Newtonian
point mass.  A next step towards GR is to modify the point mass potential to
mimic certain GR effects, such as having an innermost stable circular orbit (ISCO)
for test particle
orbits~\cite{1980A&A....88...23P,1996ApJ...461..565A}.  In many cases, a good
approximation to full GR is the conformal flatness approximation
(CFA)~\cite{1996PhRvD..54.1317W}.  CFA
assumes the spatial metric $g_{ij}$ is conformally flat at all times: 
$g_{ij}=\psi^4\delta_{ij}$.  (Throughout this paper, Latin indices run from 1
to 3, while Greek indices run from 0 to 3, with 0 being the time index.)  From
this assumption, one can derive elliptic equations for $\psi$ (from the
Hamiltonian constraint), for the shift $\beta^i$ (from the condition that
$g_{ij}$ remain conformally flat together with the momentum constraint),
and for the lapse $\alpha$ (from the maximal slicing condition).  CFA is
always accurate to first PN order; for a spherically-symmetric system it is
exact, because for such systems, there are always coordinates in which $g_{ij}$
is conformally flat.  For spinning NS, errors from assuming CFA are a few
percent~\cite{Cook:1995cp}.  The spatial metric for a spinning Kerr BH
cannot be made conformally flat (at least for a broad class of foliations)~\cite{Garat:2000pn},
and assuming CFA for spacetimes with rapidly-spinning BH has been found to introduce
significant junk radiation~\cite{Gleiser:1997ng,Hannam:2006zt}. 
  The only
way to accurately evolve general BH spacetimes is using full GR.  Another limitation
of Newtonian
and CFA calculations is that gravitational waves must be estimated using the quadrupole
approximation, while in GR simulations the waves come directly from the spacetime
evolution.

The first challenge for a full GR evolution is to acquire valid initial data. 
This must both satisfy the ADM constraint equations and represent an
astrophysically
reasonable late-inspiral binary configuration.  During the inspiral, the infall
timescale is much longer than the orbital period.  Also, gravitational radiation
will damp away any orbital eccentricity before the separation becomes small. 
The exception might be mergers in dense stellar clusters, where 3-body
interactions can be important, but these are expected to be a minority
of NSNS and BHNS binaries~\cite{2006NatPh...2..116G,2008ApJ...676.1162S}. 
(This expectation has recently been challenged, however~\cite{2009arXiv0909.2884L}.) 
Therefore, circular
orbit is usually taken to be be a good first approximation for the late inspiral.  Combining
the circular orbit assumption with CFA, one can construct quasi-equilibrium profiles of
the spacetime and the NS fluid.  By computing these profiles for many binary separations,
one can construct a sequence of ``snapshots'' that tell the history of the binary's inspiral. 
[To make each snapshot correspond to the same binary,
the baryon mass of the NS(s) and irreducible mass of the BH are fixed.]  During the late
inspiral, the quasi-equilibruim approximation breaks down and full time evolutions are needed. 
A quasi-equlibrium snapshot of the late inspiral is then used
as initial data for the evolution.  Neglecting the initial radial (i.e. infall)
velocity is known to produce spurious eccentricity in the orbit, but this
can be removed by adding back an initial infall~\cite{Pfeiffer:2007yz,
Foucart:2008qt,Kiuchi:2009jt}.  CFA produces an
initial burst of spurious ``junk'' gravitational radiation.  Junk radiation
can make it particularly difficult to create initial data with a rapidly spinning
BH.  This difficulty can be reduced by using a Kerr-Schild background metric,
rather than CFA~\cite{Lovelace:2008tw,Foucart:2008qt}.  Some junk radiation will
still be present, however, because the chosen conformal spatial metric still does not
correspond to that of a true inspiraling binary.

Evolving the metric forward in time requires choosing a formulation of GR and
choosing a gauge.  There are currently two formulations of GR used for NSNS/BHNS
simulations:  Baumgarte-Shapiro-Shibata-Nakamura
(BSSN)~\cite{Shibata:1995we,Baumgarte:1998te} and generalized
harmonic (GH)~\cite{Garfinkle:2001ni,Gundlach:2005eh,Pretorius:2004jg,Lindblom:2005qh}.  In both
formulations, the ``trick'' is to promote certain first spatial derivatives of
the metric to the status of independent variables; this turns the evolution
equations for metric components into wave-like equations.  The gauge conditions
fix the evolution of the coordinates.  For BSSN, this is done by choosing the
lapse $\alpha$ and shift vector $\beta^i$ evolution equations; the ``1+log''
lapse and ``Gamma-driver'' shift have proven immensely successful.  With these
gauges, BSSN codes can stably evolve spacetimes with BH singularities inside
the computational domain~\cite{vanMeter:2006vi,Hannam:2006vv}.  This treatment
of BH spacetimes is called the ``moving puncture'' approach because BHs can
freely move through the grid, a feature which has made it possible to do
long-term, stable evolutions of binary black hole mergers with
BSSN~\cite{2006PhRvL..96k1101C,Baker:2005vv}.  For GH, the gauge is specified through the
gauge source functions $H_{\alpha}=-g^{\beta\gamma}\Gamma_{\alpha\beta\gamma}$. 
Suitable choices for $H_{\alpha}$ are still under development
(e.g.~\cite{Pretorius:2006tp,Lindblom:2009tu}), and the moving puncture approach
has not yet been implemented in GH evolutions. 
Instead, BH singularities are excised in these
simulations, i.e. a region inside the horizon is cut out of the grid and
replaced with an inner boundary.  Pretorius has successfully simulated binary black hole
mergers using GH with excision~\cite{2005PhRvL..95l1101P,Pretorius:2006tp}. 
The Caltech-Cornell-CITA group has also successfully used GH to simulate a
variety of binary BH mergers using high-accuracy spectral techniques~\cite{Scheel:2008rj,
Szilagyi:2009qz}.

The state of the neutron star fluid at each point is given by six numbers: 
the baryon density $\rho$, temperature $T$, 3-velocity $v^i$, and electron
fraction $Y_e$.  The fluid is evolved using six conservation equations: 
baryon number conservation $(\rho u^{\mu})_{;\mu}=0$, energy-momentum
conservation $T^{\mu}{}_{\nu;\mu}=0$, and lepton number conservation
$(\rho Y_e u^{\mu})_{;\mu}=S_L$, where the lepton source term $S_L$ comes
from neutrino emission or absorption.

The stress tensor $T^{\mu}{}_{\nu}$
depends on the pressure $P$ and specific enthalpy $h$; these are determined
by the equation of state (EoS), which gives $P$ and $h$ as functions of $\rho$,
$T$, and $Y_e$.  The actual EoS of NS matter is not known, so several guesses
and approximations are used for NSNS/BHNS simulations.  (For reviews of the NS
EoS and its astrophysical implications, see~\cite{Lattimer:2000kb,Lattimer:2006xb}.) 
One is to treat the
NS as a polytrope:  $P=\kappa\rho^{1+1/n}$, where the polytropic index $n$ is
a constant.  This EoS is not temperature-dependent, so evolutions with this
EoS will not allow the NS to heat.  A more common EoS choice is to use the
polytropic law to set the initial data for a cold NS and then evolve using the
$\Gamma$-law $P=(\Gamma-1)\rho\epsilon$, where $\epsilon$ is the specific
internal energy and $\Gamma=1+1/n$.  The $\Gamma$-law is equivalent to the
polytropic law as long as the fluid is continuous and isentropic, but it will
allow heating when shocks are formed.  Polytropic evolutions are still used
to help gauge the importance of heating effects (e.g.~\cite{Baiotti:2008ra}). 
A step towards more sophisticated EoS is to break up the pressure and internal energy
into zero-temperature and thermal components, e.g. $P=P_{\rm cold}(\rho)
+P_{\rm thermal}(\rho,T)$.  Then the thermal part is set to a $\Gamma$-law
[$P_{\rm thermal}=(\Gamma-1)\rho\epsilon_{\rm thermal}$], while the cold part
can be a complicated function of $\rho$.  Recently, piecewise-polytropic laws
for the cold EoS have been proposed~\cite{Read:2008iy} and
used~\cite{Read:2009yp}.  With only four parameters, this class of EoS can
approximate any of the NS EoS proposed to date.  It thus provides a good
systematic way to cover the space of possible EoS.  Other simulations use
cold EoS based on nuclear physics calculations.  Shibata and
collaborators~\cite{Shibata:2005ss,2006PhRvD..73f4027S,Kiuchi:2009jt} have evolved NSNS
binaries in GR using
the FPS~\cite{1989nmhi.conf..103P}, SLy~\cite{2001A&A...380..151D}, and
APR~\cite{Akmal:1998cf} cold EoS.  The most realistic nuclear theory-based
EoS have a general temperature and composition dependence.  The
tabulated Lattimer-Swesty~\cite{Lattimer:1991nc} and Shen~\cite{Shen:1998gq,
1998PThPh.100.1013S} EoS have been used in Newtonian and CFA NSNS/BHNS
simulations.  Correct temperature dependence will probably be important for
modeling mergers but not inspirals.  Oechslin, Janka, and
Marek~\cite{2007A&A...467..395O} have compared many of these EoS in the context
of NSNS mergers.

The transport of energy and lepton number by neutrinos can have important
effects on the post-merger system.  A few Newtonian NSNS/BHNS merger calculations
have included these effects in approximate ways.  The simplest way is through
a leakage scheme~\cite{1996A&A...311..532R,2003MNRAS.342..673R}.  These
model neutrino cooling by removing energy from each gridpoint and adjusting
$Y_e$ at a rate set to be some function of the fluid variables and the
$\nu$-optical depth $\tau_{\nu}$.  This function is chosen to reproduce rates
given by local reaction timescales in $\nu$-transparent regions, while it
reproduces rates given by the diffusion timescale in $\nu$-opaque regions. 
Note that the scheme is local: $\nu$'s emitted in one part of the grid can't be
absorbed in another part, a fundamental limitation of leakage schemes.  Also, one
must somehow estimate $\tau_{\nu}$.  A better but more complicated neutrino
scheme, used in one study of the NSNS post-merger
system~\cite{2009ApJ...690.1681D} is flux-limited diffusion~\cite{Burrows:2006uh,
2006AAS...208.0208M}.  In this case, the neutrino radiation intensity is evolved
via a diffusion equation, with fluxes limited so that free streaming is recovered
in the $\nu$-transparent regime.  Even these simulations do not solve the full
multi-angle Boltzmann transport equation, a task beyond current numerical
resources.  Much less work has been done on $\nu$-transport in GR.  De
Villiers~\cite{2008arXiv0802.0848D} and the Illinois (UIUC)
group~\cite{Farris:2008fe} have independently developed GR radiation transport schemes
for the optically-thick limit, but these codes have not yet been applied to NSNS/BHNS binaries.

Recently, some NSNS merger simulations have included magnetic
fields~\cite{2006Sci...312..719P,Anderson:2008zp,2008PhRvD..78b4012L,
2009MNRAS.tmpL.313G}. 
These simulations all work in the ideal MHD limit--a good approximation in
most cases, given the high conductivities inside NSs.  To investigate cases
where resistive effects are important (in low-density, high-temperature regions),
Palenzuela {\it et al}~\cite{2009MNRAS.394.1727P} have developed, but not yet
applied, a code to evolve the relativistic resistive MHD equations.

Modeling the important physical effects is only one aspect of NSNS/BHNS
simulations.  There is also the computational challenge of adequately
resolving the several relevant length scales of the problem:  the gravitational
wavelength, the stellar radius, and the length scales of various fluid and MHD
instabilities.  Some simulations (e.g.~\cite{1994ApJ...432..242R,
1994PhRvD..50.6247Z,Lee:1998qk,
Rosswog:2001fh,Faber:2005yg,2007A&A...467..395O,2008ApJ...680.1326R}) use
Smoothed Particle Hydrodynamics (SPH).  The representation of the fluid by
particles can naturally adapt to changes in the fluid's size and shape.  SPH
has not yet been implemented in full GR.  Most codes use
grids to represent the fluid and metric.  The problem of multiple scales is
sometimes handled using Berger-Oliger moving-box-in-box adaptive mesh refinement
(AMR). 
This technique is used by the Carpet module of the Cactus
code~\cite{Schnetter:2003rb}, which is used in the UIUC~\cite{Etienne:2008re}
and Whisky~\cite{Giacomazzo:2007ti} GRMHD codes.  AMR is also used in the
SACRA~\cite{Yamamoto:2008js} code and by the BYU/LSU/LIU
group's code~\cite{Anderson:2006ay}.  Testing is also being done on the
use of multipatch (``cubed spheres'') grids in place of Cartesian
boxes~\cite{Zink:2007xn,2009ApJ...691..482F,Pazos:2009vb,Pollney:2009yz}. 
Multipatch grids allow one to extend the grid outer boundary $R$ at a cost
proportional to $R$ rather than $R^3$.  Finally, the Caltech-Cornell-CITA (CCC) group
uses a two-grid technique, inspired by the CoCoNuT stellar collapse
code~\cite{Dimmelmeier:2004me}.  The CCC code SpEC evolves the metric
$g_{\mu\nu}$ pseudospectrally on one grid while evolving the fluid using standard
finite-volume shock-capturing techniques on a second grid~\cite{Duez:2008rb}. 
This code's main advantage is that the fluid grid needs only extend as far
as the NS matter, rather than out to the gravitational wave zone.

A final challenge in NSNS/BHNS modeling is that of covering the entire
physically interesting range of binary parameters.  NS masses may vary by
tens of percents, while BH masses can vary widely.  In addition, the BH
can have a spin, with Kerr spin parameter $a/M_{\rm BH}$ anywhere from zero
to one, oriented in any direction.  The NS can also have a spin.  However,
the NS viscosity is thought to be too low for tidal effects to hold the NS
in corotation~\cite{1992ApJ...400..175B,1992ApJ...398..234K}, so by the late
inspiral, the orbital motion should dominate over the NS spin in most cases. 
Therefore, many simulations assume the NS to be initially irrotational.  The
effects of non-zero temperature (from tidal heating)~\cite{1992ApJ...400..175B,
Lai:1993di} and magnetic fields~\cite{2008PhRvD..78b4012L,2009MNRAS.tmpL.313G}
are expected to be negligible before the merger, and the pre-merger composition
of the NS is fixed by $\beta$-equilibrium.  Therefore, the main physical
parameters to be varied are the masses and (for BHNS binaries) the BH spin. 
Also, the EoS, since it is not known, must be treated as another parameter
(or, rather, set of parameters) to be varied.


\section{Neutron star-neutron star mergers}
\label{nsns}

\subsection{History}
\label{nsns:history}

The first numerical simulations of NSNS mergers used Newtonian physics and
polytropic EoS~\cite{1989PThPh..82..535O,1992ApJ...401..226R,1992PThPh..88.1079S,
1994ApJ...432..242R,1994PhRvD..50.6247Z,1997ApJ...490..311N}.  These studies found
that the binary merges into a massive star rotating rapidly and differentially,
but, being Newtonian, they could not check for the possibility that this object
collapses to a BH.  Simulations using CFA gravity and the zero-temperature
Mayle-Wilson EoS~\cite{1993PhR...227...97W} were carried out by
Wilson, Mathews, and Marronetti~\cite{1995PhRvL..75.4161W,1996PhRvD..54.1317W}. 
These pioneering
computations surprisingly predicted that the NSs can collapse individually to
BHs before merging.  However, they used an EoS with fairly low NS maximum mass,
and they were plagued by an error in one of their
equations~\cite{Flanagan:1998zt}, so individual NS collapse is no longer
considered likely~\cite{Shibata:1998sg}.

After this, the field bifurcated--some groups concentrated
on improving microphysics while retaining Newtonian gravity, and some groups
concentrated on improving the treatment of gravity while retaining simplified
microphysics.  In the first category would be the simulations of Ruffert, Janka,
and collaborators~\cite{1996A&A...311..532R,1997A&A...319..122R,
2003MNRAS.342..673R} and those of Rosswog and collaborators~\cite{Rosswog:2001fh,
2003MNRAS.342..673R,2003MNRAS.345.1077R}.  These simulations used the
finite-temperature EoS of Lattimer-Swesty or Shen, and they used leakage models
of neutrino cooling.  Meanwhile, binary polytropes were evolved in CFA gravity by
Oechslin, Rosswog, and Thielemann~\cite{Oechslin:2001km} and by Faber,
Grandclement, and Rasio~\cite{Faber:2003sb}.  For full GR models, an
important step was the development of accurate quasi-equilibrium configurations
to serve as initial data~\cite{Baumgarte:1997eg,Bonazzola:1998yq,Uryu:1999uu,
Gourgoulhon:2000nn,Taniguchi:2002ns}.  Finally, the first fully GR simulations of
NSNS merger were performed by Shibata and collaborators~\cite{Shibata:1999wm,
Shibata:2002jb,Shibata:2003ga}. 
Since this time, efforts have been made to combine relativistic gravity with
realistic microphysics.  Oechslin, Janka, and Marek~\cite{2007A&A...467..395O,
Oechslin:2005mw} used realistic NS EoS in CFA
simulations, while Shibata, Taniguchi, and Uryu~\cite{Shibata:2005ss,
2006PhRvD..73f4027S} modeled mergers in
full GR using realistic zero-temperature EoS (plus $\Gamma$-law thermal
components).  In the past few years, magnetic field evolution has been added
to both Newtonian~\cite{2006Sci...312..719P} and GR~\cite{Anderson:2008zp,
2008PhRvD..78b4012L,2009MNRAS.tmpL.313G} NSNS simulations.  Also, the accuracy
of these simulations has been greatly improved through the use of
AMR~\cite{Anderson:2007kz,Baiotti:2008ra,Yamamoto:2008js}. 
Finally, efforts have been made to study the effects of MHD
instabilities~\cite{Duez:2005cj,Shibata:2005mz} and neutrino energy
transport~\cite{2004MNRAS.352..753S,Lee:2005se,2006A&A...458..553S,
2009ApJ...690.1681D} on the post-merger
system.  The numerical studies to date can be combined to form a coherent picture,
our current best guess, of what happens when two neutron stars collide.

\subsection{The emerging picture}
\label{nsns:picture}

When two NS merge, they form a massive remnant.  The first question to answer
is ``does the remnant collapse to a black hole?''  It might seem like it should. 
There is a maximum mass $M_{\rm TOV,max}$ for nonrotating NS, above which the
star is dynamically unstable.  $M_{\rm TOV,max}$ depends on the unknown EoS;
predictions of its value fall in the range 1.5-2.5$M_{\odot}$~\cite{Lattimer:2000kb}. 
Two $1.4M_{\odot}$ stars will merge
into a roughly 2.8$M_{\odot}$ object.  This is significantly above even high predictions
of $M_{\rm TOV,max}$, so collapse seems likely.  However, rotation and, to a lesser
extent, shock heating increase the maximum mass.  A star uniformly
rotating at its mass-shedding limit (above which matter would be ejected from
the equator) can be stable with a mass up to $M_{\rm sup}\approx 1.2M_{\rm TOV,max}$, due
to centrifugal support~\cite{1994ApJ...422..227C}.  Rigidly rotating stars with
$M_{\rm TOV,max}<M<M_{\rm sup}$ are sometimes called supramassive.  Greater centrifugal
support, and hence greater
maximum mass, can be achieved if the star rotates differentially, with the center
having a significantly higher angular speed $\Omega$ than the
equator~\cite{1966PhRvL..17..816O,2004ApJ...610..941M}.  A NS
supported above $M_{\rm sup}$ by differential rotation is called a
hypermassive neutron star (HMNS).  Numerical experiments in GR have demonstrated that
HMNS are stable on dynamical timescales~\cite{2000ApJ...528L..29B}.  NSNS remnants
are found by simulations to have angular speeds several times higher at the
center than near the equator~\cite{Shibata:1999wm}, so they
are good HMNS candidates.

GR simulations find that, after the merger, a NSNS remnant may either collapse
promptly (i.e. on a dynamical timescale) to a BH, or it may survive as a HMNS. 
The determining factor is the mass of the NSNS system $M_{\rm NSNS}$.  If
$M_{\rm NSNS}$ is above a
threshold mass $M_{\rm th}$, the remnant collapses promptly; if
$M_{\rm NSNS}<M_{\rm th}$,
it forms a HMNS.  All of the GR codes are in agreement on this basic
point~\cite{Shibata:1999wm,Shibata:2003ga,2006PhRvD..73f4027S,Baiotti:2008ra,
Etienne:2008re}.  For $\Gamma=2$ polytropes,
$M_{\rm th}\approx 1.7 M_{\rm TOV,max}$~\cite{Shibata:2003ga}, but for more
realistic, stiffer EoS, $M_{\rm th}$ is found to be about
$1.3-1.35M_{\rm TOV,max}$~\cite{2006PhRvD..73f4027S}.  For
$M_{\rm TOV,max}=2.1M_{\odot}$, the threshold is $M_{\rm th}=2.7-2.8M_{\odot}$. 
This is a typical NSNS
binary mass, so it is quite possible that both prompt collapse and HMNS formation
occur in nature.

The prompt collapse case can be a viable GRB engine only if the post-collapse
BH is surrounded by a massive accretion disk.  For equal-mass binaries, the
disk mass is usually very small ($\ll 0.01M_{\odot}$) and the
disk is very thin~\cite{2006PhRvD..73f4027S}, making these systems poor
candidates for producing GRBs.  On the other hand, simulations find that
the merger of unequal mass binaries proceeds much differently and can produce
large
disks~\cite{1994ApJ...432..242R,Faber:2002cg,Shibata:2003ga,2006PhRvD..73f4027S,
Kiuchi:2009jt}. 
In unequal mass binaries, the lower-mass NS can fill its Roche lobe before
merger, so that it sheds matter both inward onto the more massive NS and outward
into a tidal tail.  The tail then falls back and contributes to the accretion disk
around the BH.  For a binary with mass ratio around 3:4, the disk can be
$0.01M_{\odot}$ or larger, an excellent setup for a GRB~\cite{2006PhRvD..73f4027S}. 
In addition to the mass ratio, the disk mass also depends on the mass of the
system--binaries with $M_{\rm NSNS}$ slightly above $M_{\rm th}$ produce more
massive disks than those produced by more massive binaries~\cite{Shibata:2003ga,Kiuchi:2009jt}.

For $M_{\rm NSNS}$ close to $M_{\rm th}$, the remnant survives as a HMNS,
but only for a few milliseconds.  This case was mentioned by
Shibata, Taniguchi, and Uryu~\cite{Shibata:2005ss} and studied in detail by
Baiotti, Giacomazzo, and Rezzolla~\cite{Baiotti:2008ra}.  In the latter's
simulation, the NSs merge, but then a bar-mode instability causes
the two cores to split.  The cores merge and split four times over a time of
8ms.  During this time, the remnant is highly dynamical and asymmetric, and so
it radiates gravitational waves.  This radiation carries away angular momentum,
so that the remnant loses centrifugal support and collapses, leaving a BH
surrounded by a massive ($0.07M_{\odot}$) disk.

For lower-mass systems, the product of the merger is a HMNS.  Such objects are
dynamically stable, but they are vulnerable on longer timescales to processes
(like gravitational radiation or magnetic fields) that sap the centrifugal
support in the remnant's core.  If the star is axisymmetric, its rotation will
not cause it to emit gravitational waves.  However, for rapidly rotating stars
with stiff EoS, the minimum energy configuration will be
ellipsoidal rather than spheroidal~\cite{Lai:1993ve}. (A famous example would
be the Maclaurin spheroid and Jacobi ellipsoid configurations for rapidly
rotating incompressible stars~\cite{1969efe..book.....C}.)  Stiff stars tend to
take ellipsoidal shape if $\beta=T/|W|$, the ratio of rotational kinetic to
gravitational potential energy, exceeds a critical value.  Realistic NS EoS
tend to be stiff ($\Gamma\approx 2.75$) at high densities, and numerical
simulations with Newtonian~\cite{1994ApJ...432..242R,1996PhRvD..54.7261Z},
post-Newtonian~\cite{Faber:2000uf} and full GR~\cite{Shibata:2005ss,
Baiotti:2008ra,Read:2009yp} physics have indeed found that NSNS remnants with
these EoS usually form ellipsoids.  These bar-like deviations from axisymmetry
are stronger for stiffer assumed EoS~\cite{2007A&A...467..395O}.  Such stars emit
strong gravitational waves
and lose angular momentum $J$.  This might have the effect of causing the star to
contract or of lowering the ellipticity, but simulations indicate that the
former effect dominates, and the remnant remains ellipsoidal as it contracts. 
Eventually, the remnant will reach a critical state and collapse to a BH.  From
the $J$-loss rate, the time for this to happen is inferred to be 30-100ms. 
Such long timescales are difficult to simulate in 3D with current resources, so
the first simulations stopped well before the collapse.  Oechslin and
Janka~\cite{Oechslin:2005mw} and Shibata and Taniguchi~\cite{2006PhRvD..73f4027S}
estimated the mass of the disk around the post-collapse black hole
based on the $J$ distribution in the HMNS.  Matter with specific angular
momentum high enough for orbit outside the ISCO of the (not yet formed) BH was
assumed to form the accretion disk.  These calculations give disk masses of
$\sim 0.1M_{\odot}$, high enough for GRBs.  However, these estimates necessarily
could not account for changes in the HMNS's $J$ distribution prior to collapse. 
Finally, Baiotti, Giacomazzo, and Rezzolla~\cite{Baiotti:2008ra} carried out
long-time ($\sim 30$ms) simulations using a cold polytrope EoS that, for the
first time, evolved the NSNS remnant to the point of delayed collapse (in their
case, 16ms after merger).  They found a final disk mass of
$\approx 0.08M_{\odot}$, quite close to the earlier estimates.


Prompt collapse or ellipsoidal HMNS formation occur in most simulations, but
other evolutionary paths are possible.  If $\beta$ of the remnant is below the
critical value, the HMNS may form a spheroid which will eventually be driven
to collapse by magnetic-driven, rather than gravitational wave-driven, processes
(see below).  This might happen for $M_{\rm NSNS}$ near but below $M_{\rm th}$
for some EoS~\cite{2006PhRvD..73f4027S}.  Also, if $M_{\rm NSNS}<M_{\rm sup}$,
the remnant might not collapse at all.


Shocks during the merger heat the remnant to temperatures of order 10 MeV.  At
these densities and temperatures, the remnant will radiate primarily in
neutrinos.  The HMNS has $\nu$-optical depth of $10^2-10^4$, depending on the
neutrino energy, and it will cool from radiation on a timescale of order
1~sec~\cite{2009ApJ...690.1681D}.  The neutrino emission from NSNS remnants
has been studied using leakage schemes by Ruffert, Janka, and
collaborators~\cite{1996A&A...311..532R,1997A&A...319..122R,2001A&A...380..544R}
and by Rosswog and Liebend{\"o}rfer~\cite{2003MNRAS.342..673R}.  More recently,
Dessart {\it et al}~\cite{2009ApJ...690.1681D} have modeled neutrino
transport in the HMNS under the assumption of axisymmetry
using the multi-group flux-limited-diffusion code
VULCAN/2D.  These three codes are able to extract the
neutrino signal from NSNS coalescence.  They all find that the $\bar{\nu}_e$
emission dominates over the $\nu_e$ emission, an
unsurprising result given the low $Y_e$ of the remnant.  Dessart~{\it et al}
also find high $\nu_{\mu}$/$\nu_{\tau}$ luminosities.  $\nu_i\bar{\nu}_i$
annihilation releases some $10^{49}-10^{50}$erg~s${}^{-1}$ outside the remnant. 
However, neutrino heating also drives a wind from the HMNS surface.  This wind
dumps $\sim 10^{-3}M_{\odot} {\rm s}^{-1}$ of matter into the polar regions, enough
to baryon-load the poles and prevent a GRB from being produced while the HMNS
remains~\cite{2009ApJ...690.1681D}.  Dessart {\it et al} speculate that a similar
wind may be produced by the disk after the HMNS collapses, but in this case a
centrifugal barrier might keep the polar regions baryon-clean, allowing a GRB. The
$\nu$-driven wind also carries away some $10^{-3}M_{\odot}$ of neutron-rich material
out of the HMNS system.  GR may modify the above results, given that GR
simulations tend to produce less massive tori around the HMNS than their
Newtonian equivalents, and these tori contribute disproportionately to the
neutrino emission~\cite{2003MNRAS.342..673R}.

Newtonian~\cite{1999A&A...341..499R,2001A&A...380..544R} and
CFA~\cite{2007A&A...467..395O} simulations predict that $10^{-3}-10^{-2}M_{\odot}$,
depending on the EoS, of nuclear matter is ejected during NSNS mergers.  The CFA
simulations found two components to the ejecta:  cold matter coming off the tidal
tail and hot matter ejected from the surface of contact between the stars.  As the
expelled matter decompresses, neutrons are captured onto the heavy nuclei.  The
resulting nucleosynthesis has been modeled, with various approximations concerning
the expansion rate and the initial composition and temperature, using $r$-process
network calculations~\cite{1999ApJ...525L.121F,2005NuPhA.758..587G}.  For suitable
parameters, the calculated abundances agree quite well with solar abundance pattern for
mass numbers $A>140$, especially regarding the $A=195$ peak.  However, reliable estimates
of the ejecta mass and composition require GR simulations with realistic EoS.

The effect of the magnetic field in a NSNS binary is probably negligible prior
to the merger for realistic field strengths~\cite{2008PhRvD..78b4012L,
2009MNRAS.tmpL.313G}.  For very large fields ($\ge 10^{16}$ G), magnetic
tension can suppress tidal deformation of the NSs and delay the
merger~\cite{Anderson:2008zp,2009MNRAS.tmpL.313G}.  When the stars touch, a
shear layer is formed at their interface.  The shear layer is Kelvin-Helmholtz
unstable, causing the flow to curl up into vortex rolls.  Inside these
vortex rolls, the magnetic field can be quickly amplified by orders of magnitude. 
This process was first modeled using Newtonian MHD by Price and
Rosswog~\cite{2006Sci...312..719P}.  From a $10^{12}$G pre-merger field in each
NS, they found that the field in vortex rolls reached $10^{15}$G in 1ms.  They
speculate that the field will reach equipartition in small vortices.  These
high-field pockets of matter will become buoyant, float up, and produce
relativistic blasts when they hit the NS surface.  If so, these blasts might help
provide the energy for a GRB, and magnetic pressure might help deflect the
neutrino-driven baryon wind.  Kelvin-Helmholtz-driven field amplification has
also been seen in GR simulations~\cite{Anderson:2008zp,
2009MNRAS.tmpL.313G}.  In the simulations of Giacomazzo~{\it et
al}~\cite{2009MNRAS.tmpL.313G}, the magnetic field saturates at more modest values
(e.g. a $10^{12}$G field grows by less than an order of magnitude).  The discrepancy
with the result of Price and Rosswog reflects the difficulty in numerically
capturing the smaller-scale effects introduced by the Kelvin-Helmholtz instability.

Because the angular speed of the HMNS decreases with radius, it is unstable to
the magnetorotational instability (MRI)~\cite{1998RvMP...70....1B}.  The MRI
produces small-scale
turbulence that redistributes angular momentum outward, causing the core to
slow, contract, and eventually collapse.  This process has been modeled in GR
MHD under the assumption of axisymmetry by a collaboration of the UIUC group and
Shibata~\cite{Duez:2005cj,Shibata:2005mz,Duez:2006qe}.  They found that
the HMNS collapse leaves a BH surrounded by a massive accretion disk and
baryon-clear poles.  MRI-induced delayed collapse is likely the fate of spheroidal
HMNS and ellipsoidal HMNS with very strong magnetic field.

Much of the interest in NSNS binaries is motivated by the prospect of extracting
information about NS structure from inspiral and merger gravitational waveforms. 
The inspiral waveform may be particularly important for these purposes, because
it includes the frequencies at which Advanced LIGO will be most sensitive. 
The early, low-frequency inspiral waveform can be used to constrain the NS tidal
Love number~\cite{Flanagan:2007ix}.  Faber~{\it et al}~\cite{Faber:2002zn} point out that the
energy spectrum of the late inspiral waveform can constrain the NS radius. 
Read {\it et al}~\cite{Read:2009yp} have looked for EoS signatures in the late
inspiral waveform by performing GR NSNS simulations and systematically varying the
EoS.  They conclude that the NS radius can be determined from the waveform to an accuracy
of $\sim$1 km for an event at 100 Mpc.

The merger, HMNS, and BH ringdown waveforms all have frequencies above 1kHz and so
are more difficult to detect, but the mass and EoS-related differences become quite
dramatic in them.  If $M_{\rm NSNS}>M_{\rm th}$, the inspiral signal is followed by
a merger waveform at frequecies 1kHz $\lesssim f \lesssim$ 3kHz, which in turn is
followed by a (probably undetectable) BH ringdown signal at 6.5-7kHz~\cite{Kiuchi:2009jt}. 
If $M_{\rm NSNS}>M_{\rm th}$, the spectrum has another peak at $\sim$3kHz due to the
radiation from the ellipsoidal remnant.  A HMNS lasting $\sim 50$ms located
$\sim 50$Mpc away could be detected by Advanced LIGO with a signal-to-noise ratio
of around 3~\cite{Shibata:2005xz}.  On its own, this would be a weak signal, but it
will be preceeded by a more easily-detectable inspiral signal.  The EoS strongly
affects the compaction of the HMNS, and therefore its moment of inertia and its
rotation frequency.  Therefore, the frequency of the post-merger signal peak
contains information on the EoS, particularly if $M_{\rm NSNS}$ can be extracted
from the inspiral signal~\cite{Shibata:2005xz,Oechslin:2007gn}.


\section{Black hole-neutron star mergers}

\subsection{History}
\label{bhns:history}
The first simulations of BHNS mergers were carried out a decade ago by
Lee and Kluzniak~\cite{Lee:1998qk,Lee:1999kcb,Lee:2000uz,Lee:2001ae}.  They
used Newtonian physics, with
the NS modeled as a polytrope and the BH modeled as a point mass.  Next,
simulations in Newtonian physics with nuclear-theory EoS were
performed~\cite{Janka:1999qu,Rosswog:2004zx}, and the Newtonian potential was
replaced by a Paczy{\'n}sky-Wiita potential~\cite{2005ApJ...634.1202R,
Ruffert:2009um}.  Faber~{\it et al}~\cite{Faber:2005yg} took a step toward
GR by simulating extreme mass-ratio binaries using CFA physics.  L{\"o}ffler,
Rezzolla, and Ansorg~\cite{2006PhRvD..74j4018L} modeled a head-on collision
between a black hole and a neutron star in full GR.  GR simulations of orbiting
binaries became possible after quasi-equilibrium initial data was successfully
generated~\cite{2001gr.qc.....6017M,Baumgarte:2004xq,Taniguchi:2005fr,
Taniguchi:2007xm,Grandclement:2006ht,Foucart:2008qt}.  The first full-GR
simulations of BHNS mergers were carried out by Shibata and
Uryu~\cite{Shibata:2006ks,Shibata:2006bs}, followed quickly by the
UIUC~\cite{Etienne:2007jg} and CCC~\cite{Duez:2008rb} groups.  These
simulations all used polytropic EoS and only considered the case of zero initial
BH spin.  Meanwhile, Rantsiou~{\it et al}~\cite{2008ApJ...680.1326R} modeled
BHNS mergers with spinning BH using a Kerr spacetime and an approximate
treatment for the NS self-gravity.  Their findings suggesting the importance of
BH spin were confirmed when
the UIUC group simulated the merger of several BHNS systems with nonzero initial
BH spin in full GR~\cite{Etienne:2008re}.

\subsection{The emerging picture}
\label{bhns:picture}

The mass of the black hole, $M_{\rm BH}$, can vary widely from one BHNS system
to another.  For very high $M_{\rm BH}$, the NS behaves like a test particle: 
it inspirals slowly until it reaches the ISCO
separation $d_{\rm ISCO}$, at which point it plunges into the BH and is swallowed whole. 
For $M_{\rm BH}$ of only a few times the neutron star mass $M_{\rm NS}$, the
NS will overflow its Roche lobe at a radius $d_{\rm disr}$ before a plunge takes place. 
In this case,
which resembles the unequal-mass NSNS case, the NS is tidally disrupted while
outside the BH.  Matter flows into the hole and outward into an extended tidal
tail.  Thus, we expect tidal disruption if $d_{\rm disr}>d_{\rm ISCO}$, and a
plunge into the BH otherwise.  In addition to the mass ratio, $d_{\rm disr}$
depends on the NS radius $R_{\rm NS}$.  A more compact star can survive stronger
tidal forces and will disrupt closer to the BH.  For expected $R_{\rm NS}$, the
critical mass ratio is about 4:1~\cite{Taniguchi:2007xm}.  For higher
$M_{\rm BH}$ or smaller
$R_{\rm NS}$, one would expect the NS to be swallowed whole; for lower
$M_{\rm BH}$ or larger $R_{\rm NS}$, tidal disruption and mass transfer are
expected.

The latest GR simulations by Shibata {\it et al}~\cite{Shibata:2009cn}
confirm these expectations.  Using a $\Gamma=2$ polytropic EoS and fixing
the initial BH spin to zero, they vary the mass ratio over the range
1.5:1 to 5:1, and they vary $R_{\rm NS}$ over the range 11-13km.  For mass
ratio 5:1, the NS is swallowed whole, leaving no disk.  The gravitational
waveform contains inspiral, merger, and BH ringdown signals which are similar
to a binary BH merger of the same masses.  The final kick velocity is
also similar to the binary BH case.  For mass ratios below 3:1, the
NS disrupts before falling into the hole.  The disruption spreads out the
NS matter, so that the matter is more symmetrically distributed around the BH
as it accretes, and the gravitational wave emission of the system greatly
diminishes.  This manifests itself in the Fourier spectrum of the waveform
as a steep decline in amplitude above a cutoff frequency
$f_{\rm cut}\approx 1.3f_{\rm tidal}$, where $f_{\rm tidal}$
is the wave frequency at $d_{\rm disr}$.  Since
$d_{\rm disr}$ depends on $R_{\rm NS}$, $f_{\rm cut}$ contains EoS information.  If the
star disrupts during its plunge phase, the waveform will have inspiral and merger
signals, but the ringdown will be significantly weakened.  If the the NS
disrupts during the inspiral, then the merger part of the waveform is also
suppressed.  Disruption of the NS outside the BH also results in much lower
kick velocities for the final hole than occur in the equivalent binary BH cases. 
Shibata~{\it et~al}~\cite{Shibata:2009cn}, Etienne~{\it et~al}~\cite{Etienne:2007jg},
and Duez~{\it et~al}~\cite{Duez:2008rb} all find that, under the
assumption of $\Gamma=2$ EoS and nonspinning BH, BHNS mergers usually produce
small accretion disks.  Disk masses of $\sim 10^{-2}M_{\odot}$ were obtained
by Shibata~{\it et al}~\cite{Shibata:2009cn} for very low mass ratios (e.g. 2:1)
and large NS radii ($\approx$ 14km)~\cite{Shibata:2009cn}; in all other cases, the
disk masses were much lower.  In their most recent paper~\cite{Etienne:2008re},
the UIUC group find somewhat higher disk masses:  they find a $0.06 M_{\odot}$
disk for a case with a 3:1 mass ratio, $R_{\rm NS}\approx 14$km, and no initial BH spin,
a case that had not produced a significant disk in previous studies~\cite{Etienne:2007jg,
Shibata:2009cn}.  Sources of inaccuracy in these studies include limited resolution
and inadequate treatment of low-density regions.  (See~\cite{Etienne:2008re} for a discussion
of the latter issue.)

Things are different when the BH is spinning, because $d_{\rm ISCO}$ for prograde
orbits around a spinning Kerr BH is smaller than $d_{\rm ISCO}$ for a nonspinning
hole of the same mass.  Therefore, a NS would be expected to get closer to the
BH and experience greater tidal forces before plunging. 
Etienne~{\it et al}~\cite{Etienne:2008re} have recently simulated several
BHNS systems with nonzero BH spin aligned with the orbital rotation axis.  For a mass
ratio of 3:1 and a $\Gamma=2$ EoS, they consider an aligned spin case with
$a/M_{\rm BH}=0.75$ and an anti-aligned spin case with $a/M_{\rm BH}=0.5$. 
They find that the binary with aligned BH spin merges to produce a disk
with mass $0.2M_{\odot}$ and temperature $\sim 4$MeV, a promising setup for
a GRB.  Aligned BH spin may also make tidal disruption possible in cases where
the BH is much more massive than the NS.  An affine quasi-equilibrium model by
Ferrari, Gualtieri, and Pannarale~\cite{2009arXiv0912.3692F} predicts that, for a BH
spin of $a/M_{\rm BH}=0.9$, NS will be disrupted for mass ratios up to 10:1 to 20:1,
depending on the NS mass and EoS.

The merger can be strongly affected by the NS EoS.  The EoS determines
$R_{\rm NS}$, the effects of which have already been discussed.  The EoS
also determines how $R_{\rm NS}$ changes with the NS mass, which can affect
the character of the mass transfer.  If
a small transfer of mass leads the star to expand relative to its Roche lobe,
mass transfer will be unstable.  If the star contracts into its Roche lobe,
mass transfer might be stable or turn on and off.  In general, mass transfer
tends to be more stable for stiffer EoS (higher $\Gamma$) because this corresponds
to a greater tendency for a star to shrink when it loses mass~\cite{2007RMxAC..27...57R}.

Attempts have been made to estimate the effects of NS EoS in the
context of BHNS simulations using Newtonian gravity. 
Lee and Kluzniak~\cite{Lee:1998qk,Lee:1999kcb} and
Lee~\cite{Lee:2000uz,Lee:2001ae} considered polytropes with $\Gamma$
between 5/3 and 3.  Janka,~{\it et al}~\cite{Janka:1999qu}
used the Lattimer-Swesty EoS~\cite{Lattimer:1991nc}, while Rosswog,
Speith, and Wynn~\cite{Rosswog:2004zx} used the Shen
EoS~\cite{Shen:1998gq,1998PThPh.100.1013S}.  These simulations showed
large qualitative differences for different EoS assumptions.  For
Lattimer-Swesty nuclear matter, the NS disrupts in one mass transfer event,
and a large post-merger disk is created.  For Shen nuclear matter, a NS core
can survive multiple mass transfer events, and the postmerger disk is much
smaller.  There are indications that the mass transfer is much less stable
in GR, and the differences between the cases of stiff and soft EoS is not
nearly so dramatic.  The use of
GR-mimicking potentials~\cite{1980A&A....88...23P,1996ApJ...461..565A} 
tends to eliminate episodic mass transfer~\cite{2005ApJ...634.1202R,
Ruffert:2009um}.  Simulations of large mass-ratio cases using CFA also find
generally unstable mass transfer~\cite{Faber:2005yg}.  Most recently,
the CCC group has evolved BHNS binaries with 3:1 mass ratio,
$R_{\rm NS}=14$km, and $a/M_{\rm BH}=0.5$ for a variety of EoS, including
polytropic ($\Gamma=2$ and $\Gamma=2.75$) and
Shen~\cite{our_new_paper}.  They find that the NS is disrupted in one mass
transfer event in every case, and the disk masses are all comparable
(0.1--0.2$M_{\odot}$), although stiffer EoS do produce larger, longer-lived
tidal tails.

When tidal disruption decompresses the NS matter, neutrons and protons combine
to form heavy nuclei, which heats and thickens the tidal tail~\cite{Rosswog:2004zx,
our_new_paper}.  Metzger~{\it et al}~\cite{2009arXiv0908.0530M}
point out that, if matter from the tail remains in marginally bound orbit
for around one second, even more energy may be released by r-process
nucleosynthesis, perhaps enough to unbind this matter.  Assuming a hot disk is
sometimes formed, as in the simulations of Janka~{\it et al}~\cite{Janka:1999qu},
the nucleosynthesis in an outflow from such a disk has been
calculated by Surman~{\it et al}~\cite{2008ApJ...679L.117S}.  They find that
$r$-process nucleosynthesis does occur in such outflows.  The likely BHNS merger
rate and ejecta mass make it unlikely that this is the main source of these
elements, however.


\section{The post-merger black hole plus disk system}

It appears that a massive accretion disk
($M_{\rm disk}\sim 10^{-2}-10^{-1}M_{\odot}$)
can be produced by either a NSNS or a BHNS merger.  Such disks are initially
dense ($\rho\sim 10^{10}-10^{12}$g cm${}^{-3}$), hot ($T\sim 1-10$MeV), and
neutron-rich ($Y_e\sim 0.1$).  Angular momentum transport,
most likely driven by MRI-induced turbulence~\cite{1998RvMP...70....1B},
causes matter to accrete into the black hole at enormous ``hyperaccretion''
rates of 0.1$M_{\odot}{\rm s}^{-1}$ to 10$M_{\odot}{\rm s}^{-1}$.

Steady-state models of hyperaccretion onto black holes were constructed by
Popham, Woosley, and Fryer~\cite{1999ApJ...518..356P}.  Their models were
one-dimensional (axisymmetric and vertically integrated) but incorporated
GR using a Kerr background spacetime.  Following the Shakura-Sunyaev
prescription~\cite{1973A&A....24..337S}, the angular momentum transport was
modeled by adding a shear
viscosity of strength $\nu=\alpha c_s^2/\Omega_K$, where $c_s$ is the sound
speed, $\Omega_K$ is the Keplerian angular velocity, and $\alpha$ is a
(unknown) dimensionless parameter.  Because neutrino emission was found
to be the dominant cooling process, such disks are sometimes called
``neutrino dominated advection flows'' (NDAFs).  Other steady-state, 
$\alpha$-viscosity NDAF models soon followed~\cite{2001ApJ...557..949N,
2002ApJ...577..311K,2007ApJ...657..383C}, with the most important improvement
being the inclusion of neutrino opacity effects~\cite{2002ApJ...579..706D}. 

Time evolutions of NDAFs under the influence of $\alpha$-viscosity have been
carried in 1D~\cite{2004MNRAS.355..950J}, 2D
(axisymmetry)~\cite{2002ApJ...577..893L,2004ApJ...608L...5L,
Lee:2005se} and 3D~\cite{2004MNRAS.352..753S,2006A&A...458..553S}. 
The 2D and 3D simulations used as their initial conditions the final state
of numerical BHNS merger simulations, and they treated neutrino processes
via leakage schemes.  Shibata, Sekiguchi, and
Takahashi~\cite{2007PThPh.118..257S} have taken an important step by including
magnetic fields in their NDAF evolutions.  They evolve the MHD equations in
GR but assume axisymmetry.  This allows their code to capture effects
related to the MRI, eliminating the need for an $\alpha$ viscosity.

The stresses that drive angular momentum transport also cause energy to
be dissipated as heat, raising the temperature of the disk to $T\sim 10$MeV. 
Photons in the NDAF are trapped,
so the disk cannot cool by photon radiation.  In the inner $\sim 10^3$km, the
disk is dense and hot enough to cool by neutrino emission.  (According to most
merger simulations, this would include the entire disk.)  In sufficiently
massive disks, for which the density reaches $\rho\ge 10^{11}$g cm${}^{-3}$,
neutrinos can become trapped in the inner region, and thermal energy is
advected into the hole rather than radiated.  For
$\dot M>1M_{\odot}{\rm s}^{-1}$, neutrino trapping significantly limits the
radiated energy that might power a GRB~\cite{2002ApJ...579..706D}.

To power a GRB, some $10^{50}$erg s${}^{-1}$ must be transferred from the
BH-disk system to a low-mass, ultrarelativistic outflow.  NDAF simulations
have investigated the possibility that this energy is provided by
neutrino emission.  They find that the neutrino luminosity $L_{\nu}$
increases steeply with disk mass $M_{\rm disk}$, viscous strength $\alpha$,
and corotating BH spin $a/M_{\rm BH}$~\cite{1999ApJ...518..356P,
2006A&A...458..553S,2007ApJ...657..383C,2007PThPh.118..257S}.  In favorable
circumstances ($M_{\rm disk}\sim 0.1M_{\odot}$, $\alpha\sim 0.1$,
$a/M_{\rm BH}>0.5$), $L_{\nu}\sim 10^{53}$erg s${}^{-1}$ can be released, of
which a few percent could be converted to a pair-photon fireball by
$\nu\overline{\nu}$ annihilation~\cite{2006A&A...458..553S}.  This is sufficient to
explain observed short GRB energies if the outflow is collimated into a
$\sim$1\% angle.  GRMHD simulations of NDAFs also find
$L_{\nu}\sim 10^{53}$erg s${}^{-1}$, indicating that the ``true'' effective
$\alpha$ is around 0.01-0.1~\cite{2007PThPh.118..257S}.  These simulations
also found strong variations in $L_{\nu}$ on millisecond timescales, in
agreement with the observed variability of GRBs.

\section{Future work}
\label{bhns:future}
The needed future improvements in NSNS/BHNS merger modeling fall into three
categories:  improved microphysics, improved coverage of parameter space, and
improved numerical accuracy.  For GRB modeling, accurate microphysics is the
most pressing need.  Numerical simulations aiming to determine whether or
not a sufficiently energetic, sufficiently ultrarelativistic outflow is
generated will probably need GR, MHD, and neutrino cooling, heating, and
scattering.  A full solution to the neutrino transport problem is currently
not feasible, but we may hope that approximate treatments will capture the
essential physics, such as $\nu\bar\nu$ energy deposition and $\nu$-driven
winds.  For the other main goal of providing gravitational waveform catalogs
for LIGO data analysis, covering the NSNS and BHNS parameter space is the most
important issue.  The ongoing studies of the effect of the equation of state
on the waveforms using parameterized EoS are obviously vital.  Another poorly
studied variable is the BH spin in BHNS binaries.  So far, only a few cases
have been studied in GR, and all of them belonged to the special case of
aligned spin.  Numerically, one major challenge is to improve resolution to
so that turbulent flows are adequately treated.  Another is to extend the
evolution time so that more NSNS mergers can be evolved to the point of
delayed collapse.  Fortunately, each of these issues is being pursued by at
least one research group.

\ack 
It is a pleasure to thank Joshua Faber, Francois Foucart, Yuk Tung Liu, and
this paper's referees for their suggestions.  The author is supported by a grant
from the Sherman Fairchild Foundation, by NSF grants PHY-0652952 and PHY-0652929,
and NASA grant NNX09AF96G.

\section*{References}
\bibliographystyle{iopart-num}
\bibliography{References/References}

\end{document}